\documentclass{article}

\usepackage[utf8]{inputenc} 
\usepackage[T1]{fontenc}    
\usepackage[colorlinks=true,linkcolor=black,anchorcolor=black,citecolor=black,filecolor=black,menucolor=black,runcolor=black,urlcolor=black]{hyperref}       
\usepackage{url}            
\usepackage{booktabs}       
\usepackage{amsfonts}       
\usepackage{nicefrac}       
\usepackage{microtype}     
\usepackage{xcolor}         
\usepackage{bm}         
\usepackage{graphicx}  		
\usepackage{amsmath} 
\usepackage{cite} 
\usepackage{diagbox}
\usepackage{siunitx}
\usepackage{multirow}
\usepackage{tabularx}
\usepackage{hhline}
\usepackage{arydshln}		
\usepackage{xcolor}
\usepackage{mathtools} 
\usepackage{amsthm}			
\usepackage{amssymb}			
\usepackage{algorithm}
\usepackage{pifont}
\usepackage{threeparttable}
\usepackage{fullpage}
\usepackage[letterpaper,
            left=1.67cm,
            right=1.67cm,
            top=1.8cm,
            bottom=1.69in]{geometry}

\usepackage{algpseudocode}	
\usepackage{arydshln}
\usepackage[export]{adjustbox}
\usepackage{booktabs}

\usepackage{mathtools} 
\usepackage{microtype} 
\usepackage{multirow}
\usepackage{nicefrac}       
\usepackage{pifont}   

\usepackage{url}
\usepackage{siunitx}
\usepackage{tabularx}
\usepackage{threeparttable}
\usepackage{xcolor}  

\definecolor{darkblue}{rgb}{0,0 ,0.542}
\definecolor{lightgreen}{rgb}{.9,1,.9}
\definecolor{lightred}{rgb}{1,.415,.415}
\definecolor{lightblue}{rgb}{.415,.415,1}

\newcolumntype{L}[1]{>{\raggedright\arraybackslash}p{#1}}
\newcolumntype{C}[1]{>{\centering\arraybackslash}p{#1}}
\newcolumntype{R}[1]{>{\raggedleft\arraybackslash}p{#1}}

\theoremstyle{plain} 

\def\defn{\,\coloneqq\,}

\def\C{\mathbb{C}}
\def\R{\mathbb{R}}

\def\cbm{{\bm{c}}}
\def\ebm{{\bm{e}}}

\def\xbm{{\bm{x}}}

\def\Xbm{{\bm{X}}}
\def\Ybm{{\bm{Y}}}
\def\Fbm{{\bm{F}}}
\def\Nbm{{\bm{N}}}
\def\Thetabm{{\bm{\Theta}}}

\def\Fbm{{\bm{F}}}

\def\Fbm{{\bm{F}}}

\def\Rsf{{\mathsf{R}}}

\def\argmin{\mathop{\mathsf{arg\,min}}} 

\title{TVCondNet: A Conditional Denoising Neural Network \\for NMR Spectroscopy}
\date{}

\author{Zihao Zou\textsuperscript{*}, Shirin Shoushtari\textsuperscript{*}, Jiaming Liu, Jialiang Zhang, Patrick Judge,\\
Emilia Santana, Alison Lim, Marcus Foston, and Ulugbek S.\ Kamilov
}

\begin{document}

\maketitle
\let\thefootnote\relax\footnote{\textsuperscript{*}These authors contributed equally.}
\let\thefootnote\relax\footnote{Research presented in this article is funded by the Gordon and Betty Moore Foundation  through Grant GBMF11396. (Corresponding author: Ulugbek S.\ Kamilov.)}
\let\thefootnote\relax\footnote{S. Shoushtari and J. Liu are with the Department of Electrical \& Systems Engineering, Washington University in St. Louis, St. Louis, MO 63130.}
\let\thefootnote\relax\footnote{U. S. Kamilov (email: kamilov@wustl.edu) is with the Department of Computer Science \& Engineering and the Department of Electrical \& Systems Engineering, Washington University in St. Louis, St. Louis, MO 63130.}
\let\thefootnote\relax\footnote{J. Zhang, P. Judge, E. Santana, A. Lim, and M. Foston are with the Department of Energy, Environmental \& Chemical Engineering, Washington University in St. Louis, MO 63130 USA}

\vspace{-3em} 
\begin{abstract}
\medskip\noindent
\emph{Nuclear Magnetic Resonance (NMR)} spectroscopy is a widely-used technique in the fields of bio-medicine, chemistry, and biology for the analysis of  chemicals and proteins. The signals from NMR spectroscopy often have low signal-to-noise ratio (SNR) due to acquisition noise, which poses significant challenges for subsequent analysis. Recent work has explored the potential of deep learning (DL) for NMR denoising, showing significant performance gains over traditional methods such as total variation (TV) denoising.
This paper shows that the performance of DL denoising for NMR can be further improved by combining data-driven training with traditional TV denoising. The proposed TVCondNet method outperforms both traditional TV and DL methods by including the TV solution as a condition during DL training. Our validation on experimentally collected NMR data shows the superior denoising performance and faster inference speed of TVCondNet compared to existing methods.

\end{abstract}

\section{Introduction}
\label{sec:intro}
Nuclear magnetic resonance (NMR) spectroscopy is a widely-used technique for structural determination of molecules, and  dynamics and interaction identification of macro-molecules
~\cite{inomata2009high, jardetzky2013nmr, preul1996accurate}. 
The time domain signal of NMR, known as \emph{free induction decay} (FID), has relatively high noise and thus limited NMR spectral quality~\cite{qiu2021review}. 
A common strategy to reduce the noise and enhance the SNR involves averaging multiple acquisitions. Though effective in enhancing SNR, averaging significantly extends the measurement time, thus limiting the application of NMR spectroscopy in real-time analysis~\cite{smith2015real, soong2015vivo}.

Total variation (TV) denoising~\cite{Rudin.etal1992} is a well-known technique frequently used across multiple fields~\cite{lustig2007sparse, Kamilov.etal2016}. TV has been shown to reduce noise while preserving peaks and edges in NMR spectra~\cite{joshi2015denoising}. Choosing the right regularization parameter is crucial in TV to maintain a balance between noise reduction and signal fidelity~\cite{vogel1996iterative}. Wavelet thresholding (WT)~\cite{donoho1995noising} is another widely-used denoising method used for NMR spectroscopy which is based on wavelet decomposition of the signal. The transformed signal is thresholded to reduce noise while retaining signal characteristics, requiring careful wavelet function and threshold level selection to avoid distorting the signal~\cite{altenhof2023desperate, barache1997continuous}.

FID signal is often modeled as a combination of multiple decaying exponential signals plus noise~\cite{qiu2021review, qu2015accelerated}. This modeling approach allows the FID to be restructured into a Hankel matrix, facilitating denoising through spectral analysis methods such as SVD and various matrix factorization techniques~\cite{nguyen2012denoising, qiu2021automatic, guo2023nmr}. When an NMR signal has fewer spectral peaks than the length of the FID signal, the corresponding  Hankel matrix is often considered to be low-rank. 
This insight has spurred the development of several denoising methods, including Cadzow signal enhancement~\cite{gillard2010cadzow, Condat2013,cadzow1988signal}, randomized QR decomposition (rQRd)~\cite{chiron2014efficient}, and Convex Hankel low-Rank matrix approximation (CHORD)~\cite{qiu2021automatic}.
Denoising methods such as CHORD, rQRd, and Cadzow that necessitate the decomposition of the Hankel matrix to extract singular values, are particularly time-consuming~\cite{chiron2014efficient, ying2017hankel, guo2017}.
Additionally, the quality of denoising in methods such as Cadzow and rQRd heavily relies on the estimated rank of Hankel matrix~\cite{guo2023nmr}. Thus, these methods required careful estimation of spectral peaks to derive the rank of Hankel matrix. 

\begin{figure*}[t]
    \centering
    \includegraphics[width=1\columnwidth]{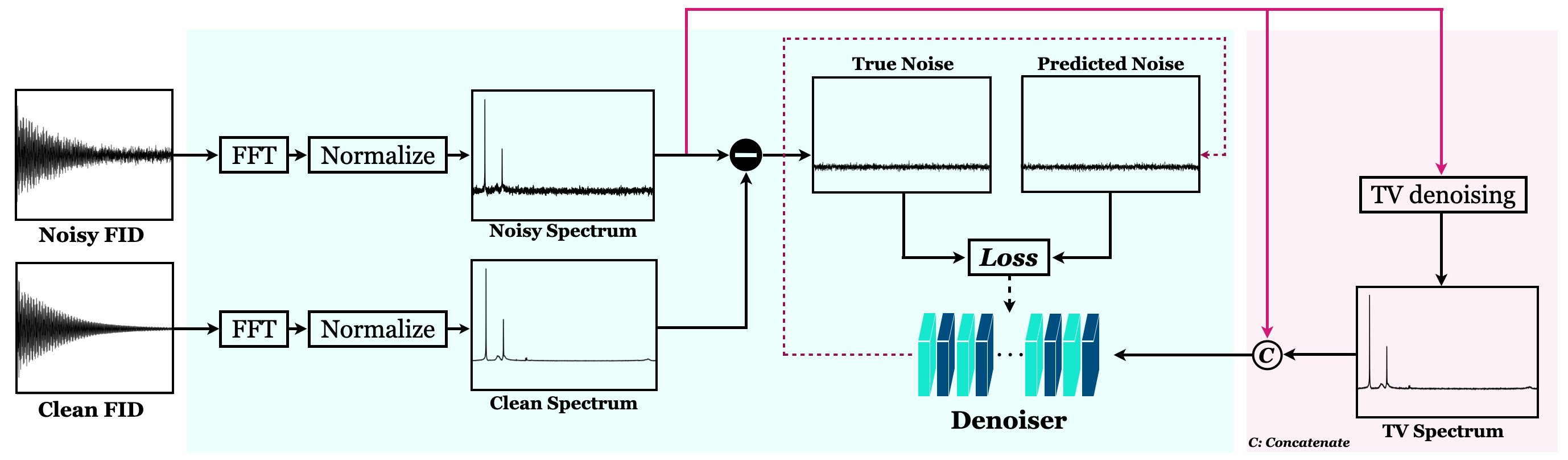}
    \caption{\small{Illustration of the \textbf{TVCondNet} pipeline.
    Noisy spectra for training are obtained by adding white Gaussian noise to the FID of the NMR signal, followed by Fourier transformation and normalization. TV denoising solution, concatenated with noisy spectra, condition the model's training to enhance its noise removal performance.
    TVCondNet is trained to predict the noise pattern from the noisy spectra with the loss calculated between the predicted and actual noise.}}
    \label{fig:pipeline}
\end{figure*}
The computational cost of SVD constrains the use of Hankel matrix decomposition in denoising large-scale NMR spectroscopy~\cite{qiu2021review}. Hence,  partial and randomized decomposition is proposed to accelerate computations by reducing the Hankel matrix size~\cite{gillard2010cadzow, chiron2014efficient}. 

The success of DL has led to the application of deep neural networks (DNNs) in spectral analysis and denoising, demonstrating superior performance over traditional NMR denoising methods. DNNs are also less computationally intensive during inference phase compared to matrix decomposition methods~\cite{chen2020review, liu2022novel, qu2020accelerated, wu2020improvement, klein2022denoising}. 
However, DNNs can result in over-smoothing of low-intensity peaks, potentially reducing peak fidelity in the denoised signals~\cite{barton2021convolution}. 

We show that the performance of DL denoising for NMR can be further improved by combining it with TV denoising. Our proposed \emph{Total Variation Conditional Denoising Neural Network (TVCondNet)} is a DNN trained by conditioning on the TV solution. The initial guess provided by TV provides an implicit regularizer for the DNN. On the other hand, data-driven training enables the refinement of the TV solution by accounting for the non-Gaussian spectral noise. Results on experimentally collected NMR spectroscopy data show that our method outperforms both traditional TV and DL solutions, by offering state-of-the-art performance and efficiency.

\section{Proposed Method}
In this section, we outline the notations and the training procedure for our proposed model, TVCondNet.

\subsection{Problem Formulation}\label{ssec:probForm}
In NMR spectroscopy denoising, the goal is to recover a clean spectrum $\Xbm^* \in \R^n $ from noisy NMR spectrum $\Ybm \in \R^n $, characterized by the following system
\begin{equation}\label{eq:noisemodel}
\Ybm = \Re(\Fbm \Nbm (\xbm + \ebm)),\quad \ebm \sim \mathcal{CN}(0, \sigma^2),
\end{equation}
where $\xbm \in \C^n$ represent the clean FID signal, $\ebm$ is complex normal noise, $\Nbm(\cdot)$ denotes normalization, and $\Fbm \in \C^{n \times n}$ is the Fourier transform.  When the clean FID of the NMR signal is accessible, clean NMR spectrum $\Xbm$ can be obtained by using the formula $\Xbm = \Re(\Fbm \Nbm (\xbm ))$. Denoising NMR spectra can be formulated as a regularized restoration problem and solved via regularized optimization
\begin{equation}
    \label{eq:inveproblem}
    \Xbm^* = \argmin_{\Xbm} \frac{1}{2}\|\Ybm - \Xbm\|^2_2 + \tau \rho(\Xbm), 
\end{equation}
where $\frac{1}{2}\|\Xbm - \Ybm\|_2$ is a data-fidelity term enforcing the consistency with the measurements, $\rho$ is a regularizer imposing desired  characteristic on the solution, and $\tau >0$ adjust regularization strength. 
TV is a widely-used regularization function that encourages  smoothness while preserving critical features like sharp peaks in the denoised spectrum~\cite{Rudin.etal1992}. The $\ell_1$-based TV regularizer, defined as  
\begin{equation}
    \label{eq:tvformula}
    \rho_{TV} (\Xbm)  \defn \sum_{i=1}^{n-1}\big |\Xbm_{i+1} - \Xbm_i\big |, 
\end{equation}
has been shown to reduce noise while preserving significant spectral details, making it suitable for spectra denoising~\cite{liao2015denoising}. 
\begin{figure*}[t]
    \centering
    \includegraphics[width=0.95\columnwidth]{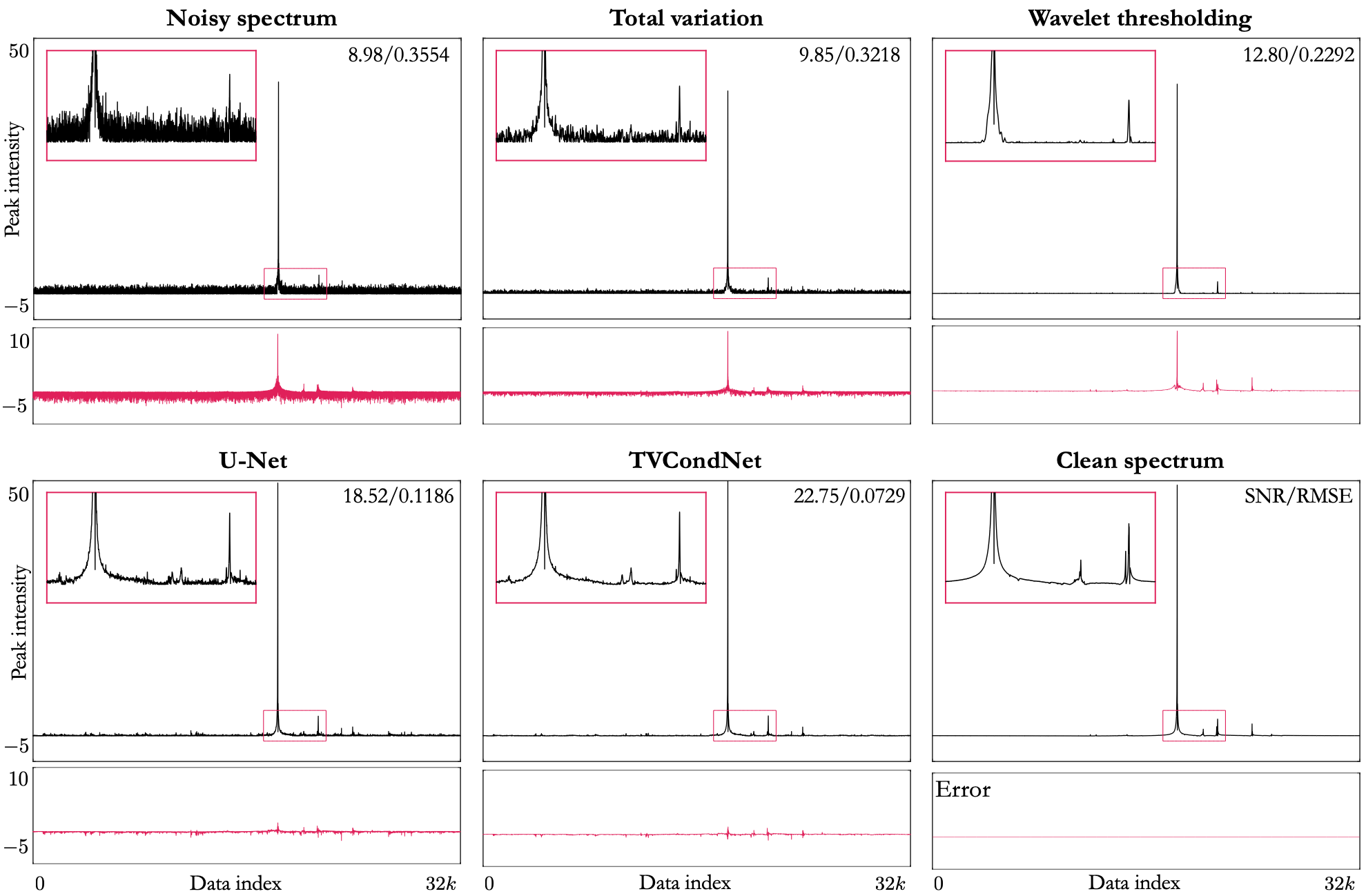}
    \caption{\small The visual comparison of NMR spectra denoising for TVCondNet and selected benchmarks. Peak intensity is visualized against data index and the performance is reported in terms of SNR and RMSE. Note the superior performance of TVCondNet in denoising, visible in the denoised spectrum (top), the zoomed-in section of the spectrum, and error visualization (bottom). }
    \label{fig:comapre}
\end{figure*}

\begin{figure*}[t]
    \centering
    \includegraphics[width=1\columnwidth]{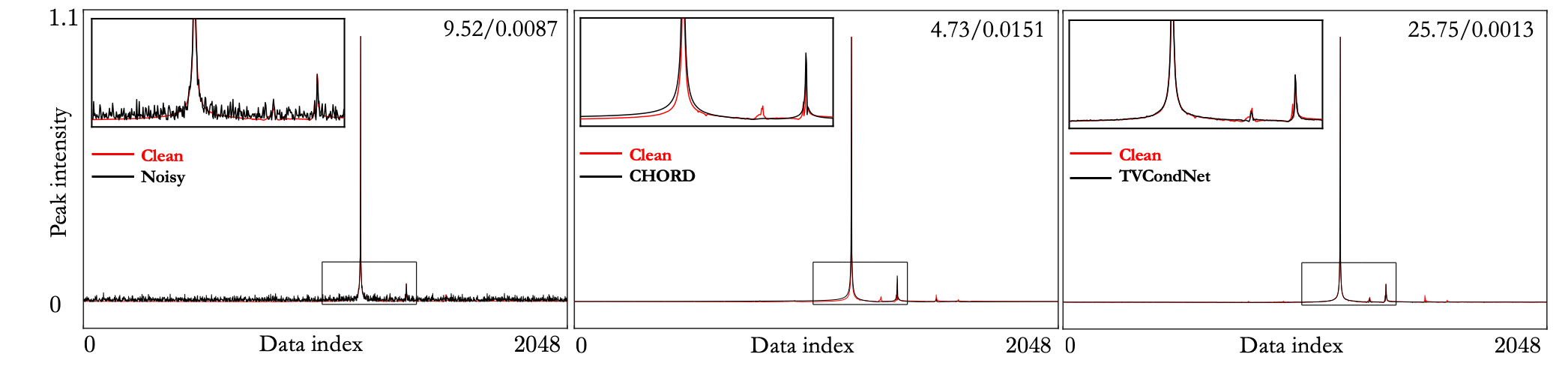}
    \caption{The visual comparison of NMR spectra denoising for TVCondNet and CHORD for cropped FID signal (2048 data points). Peak intensity is visualized against data index and the performance is reported in terms of SNR and RMSE. Note the superior denoising performance of TVCondNet.}
    \label{fig:compare2048}
\end{figure*}
\begin{table*}[t]
 \centering
 \caption{\small NMR spectra denoising for different input SNR values. Table highlights the {\color{lightblue}{\textbf{best}}} and the \textbf{second best} methods. Note the superior performance of TVCondNet for different noise levels. }
 \centering

\centering\setlength\tabcolsep{3.5pt}\renewcommand{\arraystretch}{1.2}
\begin{tabular}{l c l l c l l c l l c l l c}
\midrule
\multirow{2}{*}{\textbf{Method}} 
&& \multicolumn{2}{c}{\textbf{3 dB}}&&\multicolumn{2}{c}{\textbf{5 dB}}&& \multicolumn{2}{c}{\textbf{10 dB}}&& \multicolumn{2}{c}{\textbf{15 dB}} \\ 
\cline{3-4} \cline{6-7}\cline{9-10}\cline{12-13}
&&\textbf{SNR} $\uparrow$&\textbf{RMSE} $\downarrow$
&&\textbf{SNR} $\uparrow$&\textbf{RMSE} $\downarrow$
&&\textbf{SNR} $\uparrow$&\textbf{RMSE} $\downarrow$
&&\textbf{SNR} $\uparrow$&\textbf{RMSE} $\downarrow$\\
\midrule
\textbf{TV}    && $10.25$  &$0.321$&&$12.35 $ &$0.255$&&$ 17.87$ &$0.141$&& $23.27$& $0.078$ \\
\textbf{WT}      && $12.02 $&$0.258$ && $13.98 $ &$0.210$&&$17.53$ &$0.146$&& $19.09$&$0.125$ \\
\textbf{DnCNN}     && $16.75$  &$0.176$&& $17.68 $ &$0.152$&&$ 20.66$ &$0.103$&& $19.66$ &$0.106$\\
\textbf{U-Net}   && $\bf{21.13}$ &$\bf{0.135}$&& $\bf{22.76} $ &$\bf{0.110}$&& $\bf{25.76}$ &$\bf{0.074}$&& $\bf{30.43}$ &$\bf{0.045}$\\
\textbf{TVCondNet}                    && \color{lightblue}{$\bf{22.01}$}  &\color{lightblue}{$\bf{0.118}$}&& \color{lightblue}{$\bf{23.24}$}  &\color{lightblue}{$\bf{0.103}$}&& \color{lightblue}{$\bf{26.73}$} &\color{lightblue}{$\bf{0.065}$}&& \color{lightblue}{$\bf{30.49}$} &\color{lightblue}{$\bf{0.042}$}\\
\midrule
\end{tabular}%
\label{tab:compare}
\end{table*}

\subsection{Training procedure}
This section outlines the training procedure of TVCondNet, specifically designed for denoising NMR spectroscopy.
As depicted in Figure~\ref{fig:pipeline},  TVCondNet comprises two principal components: (1) the creation of a specialized loss function, and (2) the preparation of the conditional input for the network.

Noisy spectra $\Ybm$ are generated from the clean spectra $\Xbm$ by applying the noise model described in eq.~\eqref{eq:noisemodel}, wherein Gaussian noise is added to the clean dataset of FIDs. 
The variance of the added noise in~\eqref{eq:noisemodel} is adjusted to set the input SNR of the FID to $\{3, 5, 10, 15\}$ dB, respectively. We represent the training dataset as $\{(\Ybm_i, \Xbm_i)\}_{i=1}^K$ denoting $K$ pairs of noisy and clean training spectra . The conditional input of the neural network are derived by solving eq.~\eqref{eq:inveproblem}, incorporating the TV regularizer described in eq.~\eqref{eq:tvformula}, and are denoted as $\cbm$. The input to the denoising network is obtained by concatenating noisy spectra $\Ybm$ with the condition $\cbm$. 

For TVCondNet, we adopt the residual learning formulation to train a residual mapping $\Rsf_\Thetabm: \R^n \to \R^n$ by minimizing the following loss function
\begin{equation}
    \label{eq:lossNN}
    \ell(\Thetabm) =  \frac{1}{2K} \sum_{i=1}^K \|\Rsf_\Thetabm(\Ybm_i, \cbm_i) - (\Ybm_i - \Xbm_i)\|_2^2, 
\end{equation}
where $\Thetabm$ represents the trainable parameters of TVCondNet and $(\Ybm_i, \cbm_i)$ denotes concatenation of $\Ybm_i$ and $\cbm_i$. The denoised spectra are obtained by 
\begin{equation}
    \label{eq:solutionformula}
    \Xbm^\ast = \Ybm - \Rsf_{\Thetabm^\ast}(\Ybm,\cbm), \quad \text{s.t.}\quad \Thetabm^\ast = \argmin_\Thetabm{\ell(\Thetabm)}. 
\end{equation}
This approach enables the network to effectively learn the noise component, which is subtracted from the noisy input to yield the denoised spectra.

\section{Experimental Evaluation}
\subsection{Setup}
We obtained NMR spectra using a $500$ MHz Agilent system equipped with a OneNMR probe and DD2 spectrometer, operating on vnmrj $4.2$ software by Agilent. Over the course of several years, a total of $360$ one-dimensional ${}^1H$ and ${}^{13}C$ spectra were collected across a diverse range of samples, where $342$, $3$, and $10$ were reserved for training, validation, and testing, respectively. While slight variations in pulse sequence parameters were observed between experiments, a consistent approach was maintained, and the ${}^1H$ and ${}^{13}C$ $\pi/2$   pulses were optimized prior to each experiment.

The FID signal in NMR spectroscopy  is corrupted with Additive White Gaussian noise (AWGN) at various noise levels, corresponding to input SNR of $\{3, 5, 10, 15\}$ dB. These noisy FID signals are utilized to create four distinct datasets for training purposes. Noisy NMR spectra are generated by applying a Fourier transform to the corrupted FID signals. The resulting spectra are then normalized to ensure they have a zero mean and a standard deviation of one. The network is fed with inputs that are the concatenation of the noisy spectra and TV denoising solutions. The TV regularization parameter is fine-tuned to achieve the optimal output SNR for each training input. Consequently, the training dataset consists of pairs of noisy spectra and their corresponding TV denoising solutions, together with the clean spectra.
The test dataset is obtained by corrupting the FID signal of NMR spectroscopy with noise from the spectrometer with noise levels corresponding to input SNR of $\{3, 5, 10, 15\}$ dB. NMR noisy spectra were derived via Fourier transform and normalization, similar to training datasets. Three NMR spectra served as the validation set to fine-tune the TV denoising's regularization parameter. During testing, TVCondNet inputs were created by concatenation of the noisy spectrum with TV denoising outputs.

We chose 1D U-Net~\cite{ronneberger2015u} as a deep learning architecture to train TVCondNet.  In the training phase, we used Adam optimizer~\cite{Kingma.Ba2015} with a learning rate $1e-4$ for $100$ epochs. To address varying noise conditions, four distinct networks were trained, one for each noise level (input SNR of \{3, 5, 10, 15\} dB). The effectiveness of NMR denoising was assessed using two key metrics: Signal-to-Noise Ratio (SNR) and Root-Mean-Square Error (RMSE), ensuring a comprehensive evaluation of denoising performance
\begin{table}[t]
\caption{NMR spectra denoising for cropped FID of size $2048$. Performance (SNR(dB)) and average inference time (s) are reported for several input SNR values. Table highlights the \textbf{best} method. Note the inference time of CHORD vs. TVCondNet. }
\centering\setlength\tabcolsep{3.3pt}\renewcommand{\arraystretch}{1.2}
\scalebox{0.8}{\begin{tabular}{c c c c c c c  c c }
\midrule
\multirow{2}{*}{\textbf{Method}} 
&& \multicolumn{3}{c}{\textbf{Input SNR}} && \multicolumn{3}{c}{\textbf{Time (CPU)}}\\ 
\cline{3-5} \cline{7-9}
&&\textbf{3}  & \textbf{5}& \textbf{10} &&\textbf{3} & \textbf{5}& \textbf{10} \\
\midrule
\textbf{CHORD~\cite{qiu2021automatic}}   && {${7.96}$}   & {${8.08}$}& {${8.27}$} &&  {${145.56}$} &  {${113.20}$} & {${129.70}$}       \\
\textbf{WT~\cite{barache1997continuous}} && $19.55 $ & $20.06$& $21.07$&& $\mathbf{0.023}$  & $\mathbf{0.023}$  & $\mathbf{0.023}$  \\
\textbf{U-Net~\cite{wu2020improvement}}   && $22.97$  &$25.11$ &$27.22$ && $1.109$  & $0.754$    &$1.237 $ \\
\textbf{TVCondNet}                            && $\mathbf{24.14}$  &  $\mathbf{25.71}$ & $\mathbf{27.46}$&& $ 1.112 $  &  $0.666$   &$1.165$  \\
\midrule
\end{tabular}}
\label{tab:time}
\end{table}

\subsection{Results}
TVCondNet was benchmarked against two conventional denoising techniques—TV denoising~\cite{Rudin.etal1992} and Wavelet Thresholding~\cite{donoho1995noising}—alongside two deep learning models, DnCNN~\cite{Zhang.etal2017} and U-Net~\cite{ronneberger2015u}, and a specialized low-rank based method for NMR spectroscopy known as CHORD~\cite{qiu2021automatic}.
Given CHORD's reliance on matrix factorization, it's impractical for the large spectra of $32k$ points. For comparative analysis, FIDs in the test set were reduced to 2048 points to accommodate CHORD. Table~\ref{tab:compare} provides a performance comparison for full-length spectra, while Table~\ref{tab:time} focuses  on the cropped spectra, reporting the performance for CHORD as well. This adjustment allows us to directly compare TVCondNet's computational efficiency and performance against CHORD, highlighting TVCondNet's advantages in terms of inference time and denoising performance.
 
Table~\ref{tab:compare} shows the results of denoising methods on the test dataset. Table highlights TVCondNet's exceptional performance across various noise levels and evaluation metrics, surpassing other benchmark methods. Particularly noteworthy is TVCondNet's superior denoising performance at higher noise levels.

Figure~\ref{fig:comapre} shows the visual results of NMR spectra denoising (top) and the error of denoising (bottom) for TVCondNet and other benchmarks for input SNR of $3$ dB. The superior performance of TVCondNet  is illustrated using the zoomed-in portion of the spectra. The error visualization beneath each spectrum further emphasizes TVCondNet's potential to restore peak intensities.

Table~\ref{tab:time} presents the comparative denoising performance and inference times of TVCondNet against benchmarks for cropped FID of size $2048$. The denoising performance (left) and inference time (right) are reported for three noise levels corresponding to input SNR of $\{3,5,10\}$. The results show that TVCondNet outperforms traditional spectral denoising methods and CHORD. Notably, CHORD's significant computational time disadvantage is highlighted, underscoring TVCondNet's superior efficiency and performance. Figure~\ref{fig:compare2048} shows the visual results for denoising an NMR spectrum using CHORD and TVCondNet.

\section{Conclusion}
This work presents TVCondNet, a conditional deep learning-based denoiser for NMR spectroscopy, where TV denoising solutions are used to condition the training process of neural networks. TVCondNet shows competitive denoising performance against benchmarks, particularly excelling in recovering spectral peaks and removing  noise from experimentally collected NMR spectroscopy data. Its computational efficiency is notably superior to specialized matrix factorization methods designed for NMR denoising such as CHORD.

\bibliographystyle{IEEEbib}
\bibliography{refs}
\end{document}